\documentclass[%
pra, reprint,
superscriptaddress,
 amsmath,amssymb,
 aps,
 nofootinbib,
]{revtex4-1}
\usepackage{amssymb}
\usepackage{hhline}
\usepackage{amsmath}
\usepackage{graphicx}
\usepackage{dcolumn}
\usepackage{bm}
\usepackage{hyperref}
\usepackage{subfigure}
\newcolumntype{P}[1]{>{\centering\arraybackslash}p{#1}}
\newcolumntype{M}[1]{>{\centering\arraybackslash}m{#1}}
\newcommand\ddfrac[2]{\frac{\displaystyle #1}{\displaystyle #2}}

\begin{document}
\title{Quantum absorption refrigerator with trapped ions}
\author{Gleb Maslennikov}
\altaffiliation{These authors contributed equally to this work}
\author{Shiqian Ding$^\dagger$}
\altaffiliation{These authors contributed equally to this work}
\author{Roland Habl{\"u}tzel}
\author{Jaren Gan}
\author{Alexandre Roulet}
\author{Stefan Nimmrichter}
\author{Jibo Dai}
\affiliation{Centre for Quantum Technologies, National University of Singapore, 3 Science Dr 2, 117543, Singapore}
\author{Valerio Scarani}
\author{Dzmitry Matsukevich}
\affiliation{Centre for Quantum Technologies, National University of Singapore, 3 Science Dr 2, 117543, Singapore}
\affiliation{Department of Physics, National University of Singapore, 2 Science Dr 3, 117551, Singapore}

\renewcommand*{\thefootnote}{\fnsymbol{footnote}}
\setcounter{footnote}{1}
\footnotetext{Present address: JILA, National Institute of Standards and Technology and University of Colorado, and Department of Physics, University of Colorado, Boulder, CO 80309, USA }
\renewcommand*{\thefootnote}{\arabic{footnote}}
\setcounter{footnote}{0}

\begin{abstract}
Thermodynamics is one of the oldest and well-established branches of physics that sets boundaries to what can possibly be achieved in macroscopic systems.
While it started as a purely classical theory, it was realized in the early days of quantum mechanics that large quantum devices, such as masers or lasers, can be treated with the thermodynamic formalism~\cite{GJ1959,geusic_1967}. 
Remarkable progress has been made recently in the miniaturization of heat engines~\cite{steeneken_piezoresistive_2011} all the way to the single Brownian particle~\cite{MI2016,krishnamurthy_bacterial_2016} as well as to a single atom~\cite{RJ2015}. 
However, despite several theoretical proposals~\cite{RJ2013,CY2012,brunner_refrigerator,VD2013}, the implementation of heat machines in the fully quantum regime remains a challenge. 
Here, we report an experimental realization of a quantum absorption refrigerator in a system of three trapped ions, with three of its normal modes of motion coupled by a trilinear Hamiltonian such that heat transfer between two modes refrigerates the third. 
We investigate the dynamics and steady-state properties of the refrigerator and compare its cooling capability when only thermal states are involved to the case when squeezing is employed as a quantum resource. 
We also study the performance of such a refrigerator in the single shot regime~\cite{MM2015}, and demonstrate cooling below both the steady-state energy and the benchmark predicted by the classical thermodynamics treatment.
\end{abstract}
\maketitle

\begin{figure}[ht!]
\centering
\includegraphics[width=0.95\columnwidth]{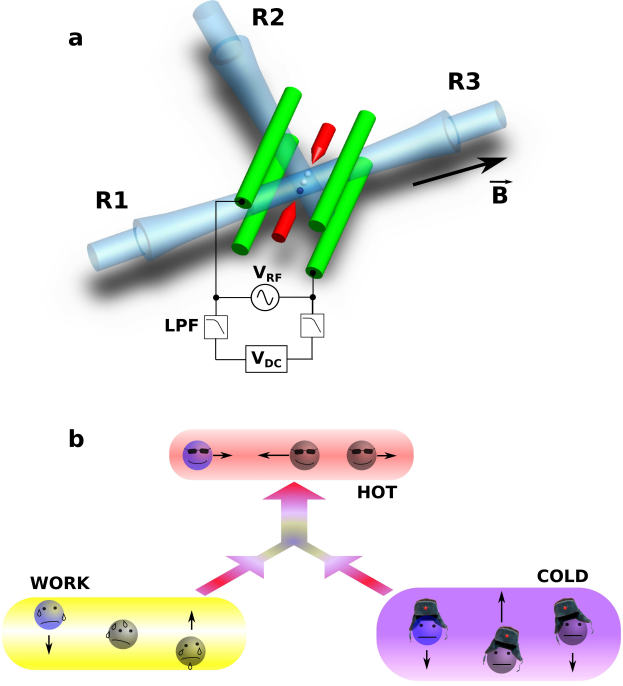}
\caption{\label{fig:heat_flow}
{\bf Experimental setup. a}. Schematic of the linear rf-Paul trap with three trapped {$^{171}$Yb$^+$} ions. The Raman beams (R1, R2, and R3) are responsible for applying the optical dipole force for the state preparation, and for coupling the ions motional modes to the internal state during the motional state detection. Two (grey) ions are prepared in the $^{2}F_{7/2}$ ``dark" state (see Methods). The radial confinement of the ions provided by radiofrequency (RF) potential, can be fine tuned by adjusting the offset voltage applied to the diagonally opposite trap electrodes. The speed of this tuning is controlled by a pair of low-pass filters (LPFs). {\bf b,} Direction of heat flow in the absorption refrigerator. Energy from the work body is transferred to the hot body, which pumps energy from cold body to hot body. The black arrows label the motional eigenmodes utilized as heat bodies}
\end{figure}

Rapid progress in the experimental control of small quantum systems revives interest in the merging of thermodynamics with quantum mechanics~\cite{an_jarzynski_2015,Kaufman794,2016ergodicNP,2016MBLNP} and poses fundamental questions: 
What is the smallest heat machine one can build~\cite{LN2010}? Can quantum effects improve the performance of a heat engine, and if so, can we use quantum correlations as a fuel~\cite{SM2003,JP2001,RK2014,GE1992}? While a lot of work in this 
field is focused on heat engines, we consider here another standard example of a heat machine: the absorption refrigerator. 
The first such device was invented in 1850 by the Carr\'e brothers~\cite{carre_fridge_1860}  and was one of the first practical refrigerators used in industry.
Modern designs incorporating numerous technical improvements~\cite{einstein_fridge_1930} remain a popular choice of refrigeration devices~\cite{CY2006}.
In general (Fig.~\ref{fig:heat_flow}), an absorption refrigerator consists of three parts: cold, hot and work bodies.
It makes use of heat from the work body to cool down the cold one, while transferring heat to the hot body. 
Although the classical thermodynamics of the absorption refrigerator is well understood~\cite{CY2006}, its description in terms of quantum mechanics is still a subject of numerous theoretical studies~\cite{LN2010,LA2012,RK2014,CL2014}, and several proposals to implement it in the quantum regime using a system of superconducting qubits~\cite{CY2012,brunner_refrigerator} or quantum dots~\cite{VD2013} exist in the literature.

Here we study the performance of absorption refrigeration in the quantum regime, utilizing the modes of motion of trapped Ytterbium ions as the heat bodies (Fig.~\ref{fig:heat_flow}). We experimentally test two hypotheses: 
whether the absorption refrigerator performs better when the thermal state of the work mode is squeezed~\cite{CL2014}, and whether there is an advantage when operating in the single-shot cooling regime~\cite{MM2015}.
The latter relies on coherent population oscillations that can occur among the coupled modes in the quantum system before the steady state is reached.

The interaction Hamiltonian in the system of three ions, induced by anharmonicity of the Coulomb repulsion between the ions, has the form~\cite{james_complete_2003,LA2012}
\begin{equation}
\label{eq:interaction}
\hat{H} =  \hbar \xi ( \hat{a}_h^{\dagger} \hat{a}_w \hat{a}_c + \hat{a}_h \hat{a}_w^{\dagger} \hat{a}_c^{\dagger} ),
\end{equation}
where the $\hat{a}_i$ ($\hat{a}_i^\dagger$) are the annihilation (creation) operators for the corresponding harmonic oscillators labeled by $i=h,w,c$, and $\xi = 9 \omega_z^2\sqrt{\hbar/m \omega_h \omega_w \omega_c} / 5 x_0$ is the coupling rate. 
Here $x_0 = (5 e^2 / 16 \pi \epsilon_0 m \omega_z^2)^{1/3}$ is the equilibrium distance between the ions, $m$ is the ion mass, $e$ is the ion charge, $\epsilon_0$ is the vacuum permittivity, and $\omega_z$ is the single ion axial trap frequency. The Hamiltonian~\eqref{eq:interaction} is valid in the rotating wave approximation when the mode frequencies satisfy the resonance condition $\omega_h = \omega_w + \omega_c$.

To understand how this absorption refrigerator works~\cite{LA2012}, consider the following scenario:  
When the temperature of the work ($w$) mode is higher than the hot ($h$) mode, some energy tends to flow from the former to the latter.
Due to the structure of the interaction Hamiltonian~\eqref{eq:interaction}, transfer of energy from ($w$) to ($h$) is always accompanied by energy transfer from ($c$) to ($h$) resulting in the cooling of the cold ($c$) mode.
At some temperatures this process is balanced by the flow of the energy in opposite direction, leading to an equilibrium.
For thermal states, it then requires the mean phonon numbers $\bar{n}_i^{(eq)}$ to fulfill (see Methods)
\begin{equation}
\label{eq:second_law_nbar_text}
\left(1 + \frac{1}{\bar{n}_h^{(eq)}}\right)  = \left(1 + \frac{1}{\bar{n}_w^{(eq)}}\right) \left(1 + \frac{1}{\bar{n}_c^{(eq)}}\right).
\end{equation}
If the system is initially prepared at $(\bar{n}_h^{(in)},\bar{n}_w^{(in)},\bar{n}_c^{(in)})$ away from equilibrium, the interaction Hamiltonian~\eqref{eq:interaction} can only lead to mean phonon numbers $(\bar{n}_h^{(in)}-\epsilon_h,\bar{n}_w^{(in)}+\epsilon_w,\bar{n}_c^{(in)}+\epsilon_c)$ such that $\epsilon_c = \epsilon_w = \epsilon_h$.

\begin{figure}
\centering
\includegraphics[width=1.00\columnwidth]{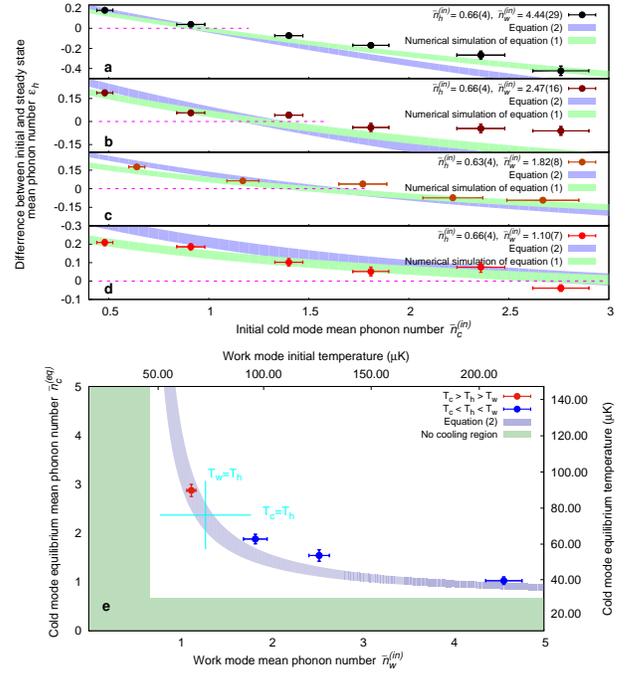}
\caption{
\label{fig:axial_mode_evol_and_equil}
{\bf Absorption refrigeration demonstration. (a-d),} The difference $\epsilon_h = \bar{n}_h^{(in)} - \bar{n}_h^{(ss)}$ plotted against the initial cold mode phonon number, $\bar{n}_c^{(in)}$ for different initial $\bar{n}_w^{(in)}$. The shaded curves are predictions of equation~\eqref{eq:second_law_nbar_text} (blue) and numerical simulations of~\eqref{eq:interaction} (green), taking experimental uncertainty of initial state preparation into account. The numerical simulations of~\eqref{eq:interaction} agree well with the experiment.
The equilibrium cold mode phonon number $\bar{n}_c^{(eq)}$, which corresponds to $\epsilon_h = 0$ (dashed line), is determined by a linear extrapolation using two points that are closest to it. {\bf e,} The obtained values are then plotted against initial $\bar{n}_w^{(in)}$ and compared to the predictions of equation~\ref{eq:second_law_nbar_text}. The absorption refrigeration occurs at the region at which the cold mode temperature is the lowest (blue dots). The red dot does not reach the refrigeration effect since work mode temperature is lower. The green zone indicates no cooling according to inequality~\eqref{eq:minWorkOcc}.}
\end{figure}

To demonstrate the equilibrium performance of the refrigerator, we start with all modes prepared in thermal states (see Methods). We fix $\bar{n}_h^{(in)}$ in the hot mode and change $\bar{n}_w^{(in)}$ and $\bar{n}_c^{(in)}$.
For each value of $\bar{n}_w^{(in)}$ and $\bar{n}_c^{(in)}$ we measure the mean phonon numbers of the hot mode at long interaction times $\tau\gg\xi^{-1}$ at which they have effectively converged to their long-time average values (see Fig.~\ref{fig:axial_mode_evol_and_equil}).
For each  $\bar{n}_w^{(in)}$  we find an equilibrium $\bar{n}_c^{(eq)}$ which gives $\epsilon_h=0$ and plot these points in panel (e). For majority of the points shown on the Fig.~\ref{fig:axial_mode_evol_and_equil} the temperatures $T_i=\hbar \omega_i[k_B\mathrm{ln}(1+1/n_i)]^{-1}$ corresponding to the mean phonon numbers $n_i$, satisfy the  condition $T_{c}^{(eq)}<T_{h}^{(eq)}<T_{w}^{(eq)}$ that implies the refrigeration of the cold mode~\cite{CY2006,LA2012}.  Here $k_B$ is the Boltzmann constant.

We further notice in Fig.~\ref{fig:axial_mode_evol_and_equil}\,(a-d) that experimental points systematically disagree with Equation~\eqref{eq:second_law_nbar_text} away from equilibrium. Indeed, the numerical simulations of~\eqref{eq:interaction} (see Methods) predict that the system approaches a non-thermal and correlated steady state in the long-time limit. 
The steady state of each mode is then better characterized by the mean phonon number $\bar{n}_i^{(ss)}$ (energy) rather than temperature. For cooling the cold mode ($\epsilon_c < 0$), the following inequality must be satisfied~\cite{CL2014, LA2012} (See Methods)
\begin{equation}
\label{eq:minWorkOcc}
 \bar{n}_w^{(in)} > \bar{n}_h^{(in)}  \frac{1 + \bar{n}_c^{(in)}}{\bar{n}_c^{(in)} - \bar{n}_h^{(in)}}.
\end{equation}
To investigate the cooling properties away from equilibrium, we focus on the mean phonon number of the cold mode whose temporal evolution is shown in Fig.~\ref{fig:cooling_cold_mode}\,(a-f).
For $\bar{n}_h^{(in)} = 0.66(4)$ and $\bar{n}_c^{(in)} = 2.63(13)$, {we observe a nett decrease of the cold mode mean phonon number in panels (a,b), equilibrium in (c), and an increase in (d-f). The data points are plotted relative to the computed long-time averages and show good agreement with theory. The nett difference of the final steady state from the initial mean phonon number is shown in Fig.~\ref{fig:cooling_cold_mode}\,g. Again the long-time average values predicted by quantum theory match the experiment quite well, while prediction  of equation~\eqref{eq:second_law_nbar_text} disagrees with the data.

\begin{figure*}
\centering
\includegraphics[width=1.70\columnwidth]{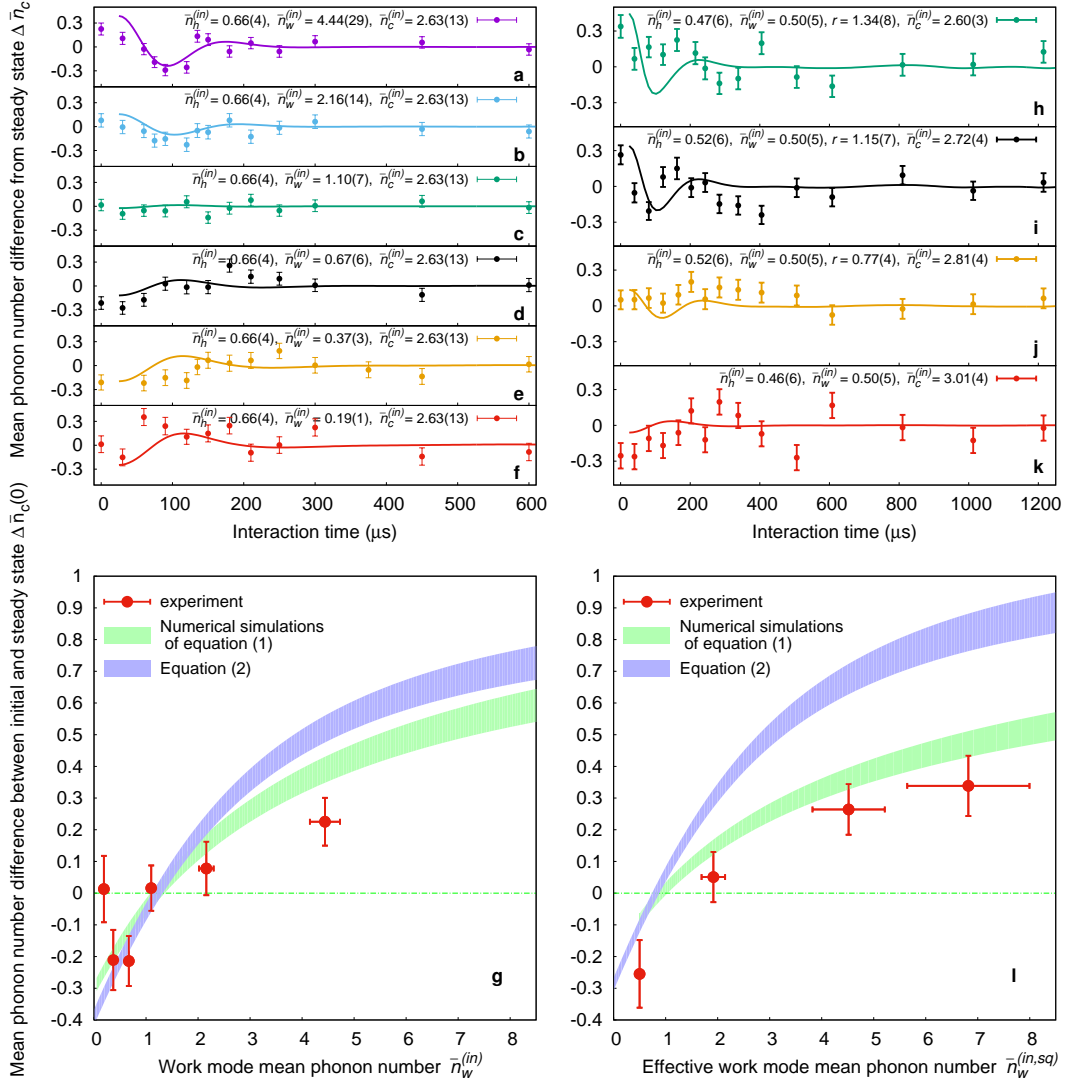}
\caption{\label{fig:cooling_cold_mode}
{\bf Non-equilibrium evolution of the cold mode with and without work mode squeezing. (Top),} The difference $\Delta\bar{n}_c(\tau) = \bar{n}_c(\tau) - \bar{n}_c^{(ss)}$ of the measured values $\bar{n}_c(\tau)$ and steady state value $\bar{n}_c^{(ss)}$ is shown as a function of $\bar{n}_{w}^{(in)}$ for purely thermal ({\bf a-f}) and squeezed thermal states ({\bf h-k}) initially prepared in the work mode. The solid lines are numerical simulations of the state evolution using experimental initial conditions. {\bf (Bottom),} The difference between the steady state and the initial mean phonon numbers in the cold mode $\Delta\bar{n}_c(0) = \bar{n}_c^{(in)} - \bar{n}_c^{(ss)}$ plotted against $\bar{n}_w^{(in)}$ for thermal ({\bf g}) and squeezed thermal ({\bf l}) states of the work mode. The blue shaded curves represent equation~\eqref{eq:second_law_nbar_text} predictions, while the green shaded curves are numerical simulations of the state evolution under Hamiltonian~\eqref{eq:interaction}. Both curves take into account the experimental uncertainty of initial state preparation.
A slight difference in initial hot mode values does not change the gradient of the $\bar{n}_c^{(ss)}$ dependence on the $\bar{n}_w^{(in)}$ values, but only results in a shift along the vertical axis.}
\end{figure*}
 
We next study the influence of quantum mechanical coherence. We prepare a squeezed thermal state of the work mode~\cite{CL2014} and compare cooling performance to the case where the mode is prepared in a thermal state of the same mean phonon number.
Squeezing increases the mean phonon number from $\bar{n}_w^{(in)}$ to  $\bar{n}_w^{(in,sq)}(r)=\bar{n}_w^{(in)}$cosh$(2r)~+$~sinh$^2(r)$, with $r$ the squeezing parameter~\cite{1989squeezedthermalKnight}. 
The experiment is repeated for several values of $r$, keeping $\bar{n}_w^{(in)}$ fixed.
When $r$ increases (Fig.~\ref{fig:cooling_cold_mode}(h-k)}), the system undergoes a transition from heating to cooling, as can be seen from the evolution of the cold mode.
We plot the difference of final and initial mean phonon numbers in panel (l) for direct comparison to the previously discussed thermal case in (g). The nett change in the mean phonon number is now smaller, which implies that squeezing of the work mode decreases the cooling performance. Further simulations show that, for a fixed $\bar{n}_w^{(in)}$, cooling is indeed most effective when no squeezing is applied at all.

\begin{figure}
\centering
\includegraphics[width=1.0\columnwidth]{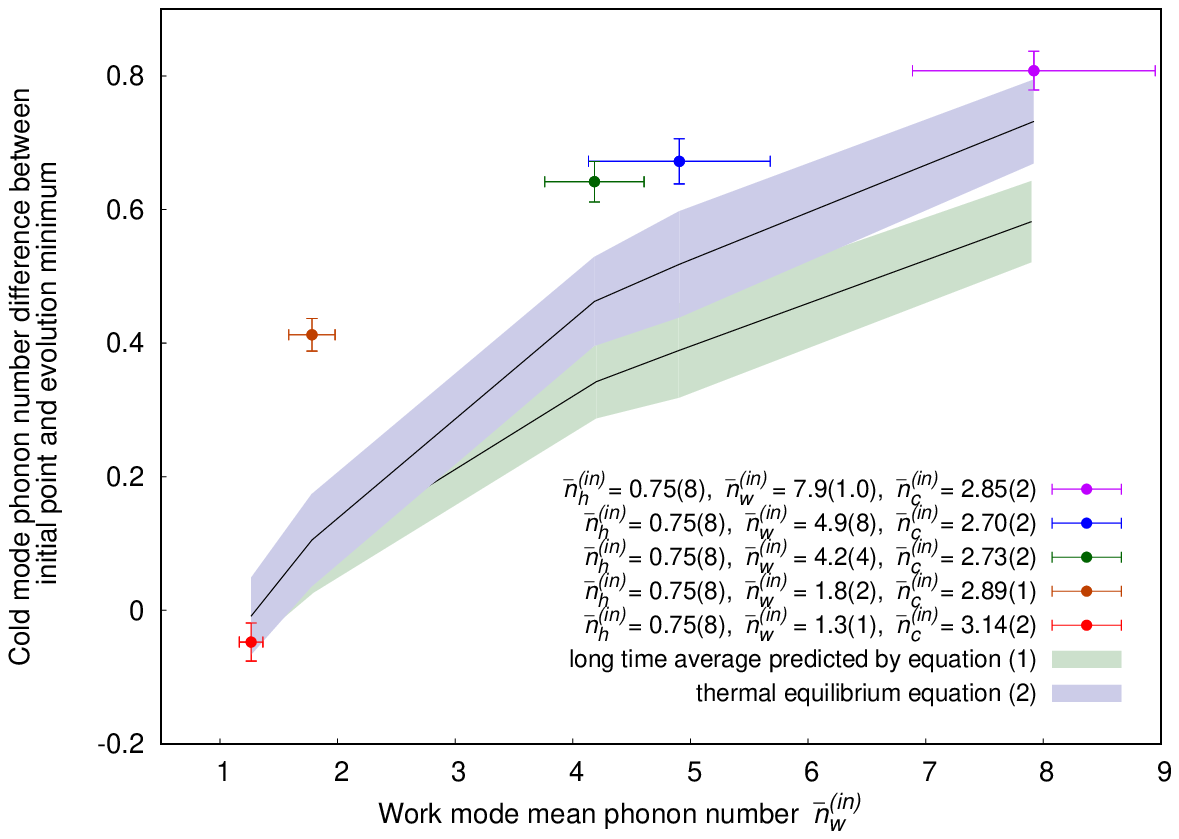}
\caption{\label{fig:eq_violation}
{\bf Absorption refrigerator operating in the single shot regime.} The difference $\bar{n}_c^{(in)}-\bar{n}_c(\tau)$ between the measured initial phonon number and the mean phonon number at interaction time $\tau$ that gives the strongest cooling (colored points), is shown for several $\bar{n}_w^{(in)}$. The blue shaded region corresponds to the range of predicted by~\eqref{eq:second_law_nbar_text} after taking the experimental uncertainty of $\bar{n}_i^{(in)}$ into account and the green shaded region is the long time average predicted by numerical simulations. The uncertainty in the x-axis is the error of the fit to the measured initial work population, while the uncertainty in the y-axis represents one standard deviation of the statistical uncertainty.}
\end{figure}

We now focus on the single-shot cooling method~\cite{MM2015}. Here the interaction is switched off at the right moment such that the evolution halts at a transient state with a lower mean phonon number $\bar{n}_c$ than the steady state one $\bar{n}_c^{(ss)}$. 
Conversely, one would achieve a higher mean phonon number than the steady-state value in the heating regime. Both regimes can be seen in Fig.~\ref{fig:cooling_cold_mode}\,(a-f), where the greatest deviation of $\bar{n}_c$ from the initial $\bar{n}_c^{(in)}$ is consistently reached at about $100\,\mu$s of interaction time, $\tau\approx(2\xi)^{-1}$.
We plot in Fig.~\ref{fig:eq_violation} the difference between the initial and this value for different work mode phonon numbers $\bar{n}_w^{(in)}$. The first data point at $\bar{n}_w^{(in)}=1.3(1)$ has a vanishing difference since it corresponds to the system at thermal equilibrium according to~\eqref{eq:second_law_nbar_text}. However, the difference increases with growing $\bar{n}_w^{(in)}$, and exceeds the long-time steady-state value consistently and significantly---a striking demonstration of the advantage of coherence-assisted single-shot cooling~\cite{MM2015}. 
Note that the cooling even exceeds the thermal equilibrium values set by~\eqref{eq:second_law_nbar_text}, which could be viewed as a classical thermodynamics benchmark. 
The ability to cool more efficiently and on a shorter timescale is related to the transient coherence generated during the unitary time evolution under the trilinear Hamiltonian~\eqref{eq:interaction}. Indeed, it is easily shown that an incoherent version of the trilinear interaction (see Methods) precludes the device from cooling further than the steady-state occupation~\cite{MM2015}. 

In conclusion, we have demonstrated an implementation of an absorption refrigerator utilizing the harmonic modes of motion in a trapped-ion system. We have shown that the classical concept of the absorption refrigerator can be extended to the quantum domain, and its cooling power per unit mass ($\hbar \omega_c \Delta n_c/(3\,m\,\tau) \approx$ 2.4 W/kg) is comparable to its classical counterparts \footnote{assuming a fridge compressor with a cooling capacity 140W and mass 7.5kg (model ZEL HPZ100A), the cooling power per unit mass is $\approx ~$19 W/kg.}. 
The experiment confirms our theoretical understanding of the refrigerator dynamics and its steady-state characteristics based on a coherent three-body interaction model. In particular, we could observe that, starting away from equilibrium, the system energies rapidly approach a steady state, even in the absence of environmental coupling. Simple arguments based on equilibrium thermodynamics do not predict this steady state, although they give the correct temperature requirements \eqref{eq:minWorkOcc} for cooling.
While it was shown that utilizing squeezed states allows the refrigerator to transition from a heating to cooling regime, hence demonstrating that squeezing could be used as a quantum fuel, when directly compared to the thermal work reservoir we could also observe a diminished performance of the refrigerator. This leads to the surprising implication that exploiting quantum resources does not necessarily enhance, but may even be detrimental to the performance of heat machines---an issue worth studying further.
On the other hand, we demonstrate a significant advantage with respect to both steady-state cooling and thermodynamic benchmark when we operate the refrigerator in a single-shot regime. 

\section*{Methods}
\subsection{Steady state populations} 

To gain insights on the operation of the absorption refrigerator we first consider an ideal adiabatic process that satisfies~\cite{RK2014,CL2014}
\begin{equation}
\label{eq:second_law}
 \Delta \dot{S} = \frac{\dot{Q}_h}{T_h} + \frac{\dot{Q}_w}{T_w} + \frac{\dot{Q}_c}{T_c} = 0.
\end{equation}
Here $\dot{Q_i} = \hbar \omega_i \dot{n}_i$ is the energy per unit time flowing to the mode $i$ from its bath at temperature $T_i$. 
Using the Bose-Einstein distribution with mean phonon number $\bar{n}_i$ for each mode,
$1/T_i = \frac{k_B}{\hbar \omega_i} \ln (1 + 1/\bar{n}_i)$, and the constraint $\dot{n}_h = - \dot{n}_w = - \dot{n}_c$ implied by the Hamiltonian~\eqref{eq:interaction}, equation~\eqref{eq:second_law} reduces to \eqref{eq:second_law_nbar_text}. The cooling condition~\eqref{eq:minWorkOcc} is obtained by noting that during cooling $\dot{n}_c < 0$ and $\Delta \dot{S}> 0$ (Second Law of Thermodynamics).

The corresponding quantum state $\rho = \rho_h \otimes \rho_w \otimes \rho_c$ is stationary as it commutes with the interaction Hamiltonian~\eqref{eq:interaction}. Conversely, if the system is prepared out of equilibrium the trilinear interaction cannot drive it towards another equilibrium of this type. We nevertheless observe (Fig.~\ref{fig:cooling_cold_mode}) that the unitary time evolution after long interaction time leads to an~\textit{effective} equilibration~\cite{short_and_farreley_2012,eisert:2015} of the mode energies at values corresponding to the infinite time average of the system state, $\rho_\infty =  \lim_{t\to \infty} \frac{1}{t} \int_0^t {\rm d} \tau \,\rho(\tau)$. This asymptotic state, which is also obtained by dephasing the initial ensemble in the eigenbasis of~\eqref{eq:interaction}, is not thermal and carries correlations between the three modes.

Note that by resorting to the unitary evolution of initially prepared thermal states, we are considering the regime of fast internal refrigerator dynamics and slow thermalization. High thermalization rates of the order of the coupling frequency $\xi$ would only thwart the coherent dynamics required for single-shot cooling.

\subsection{Experimental setup}
The detailed description of our setup can be found elsewhere~\cite{
ourpaper_parametricoscillator}. In brief, we trap three $^{171}$Yb$^{+}$ ions in a linear rf-Paul trap (see Fig.~\ref{fig:heat_flow}\,a). The single ion trap frequencies are $(\omega_x,\omega_y,\omega_z) = 2\pi\times(1025.1,937.7,570)~\mathrm{kHz}$ for the data presented in Fig.~\ref{fig:axial_mode_evol_and_equil} and Fig.~\ref{fig:cooling_cold_mode}\,(a-g), and $(\omega_x,\omega_y,\omega_z) = 2\pi\times(764.9,701.8,425.3)~\mathrm{kHz}$ for Fig~\ref{fig:cooling_cold_mode}\,(h-l) and Fig.~\ref{fig:eq_violation}.
The radial frequencies are actively stabilized (drift $<200~\mathrm{Hz/hour}$) and can be fine tuned by DC offset voltages applied to two diagonally opposite trap electrodes, while the axial frequency is fixed and has negligible systematic drift.  The normal modes chosen to represent the hot, work, and cold bodies are the axial zigzag, the radial rocking, and the radial zigzag mode (Fig.~\ref{fig:heat_flow}\,b), with frequencies $\omega_h = \sqrt{29/5}~\omega_z$, $\omega_w = \sqrt{\omega_x^2 - \omega_z^2}$, and $\omega_c = \sqrt{\omega_x^2 - 12~\omega_z^2 / 5}$, respectively. The measured coherence time of a single phonon in all the modes ($\geq 8\,\mathrm{ms}$) is much larger than the time required to achieve the steady state. 

A frequency-doubled, mode-locked Ti:Sapphire laser generates 250 mW at central wavelength of 374 nm with pulse width of 3 ps and repetition rate of 76.2 MHz, and is used to achieve spin-motion coupling~\cite{Hayes_2010} and to apply the optical dipole force to the ion~\cite{ourpaper_microwave}. At all times two of the three trapped ions are pumped into a dark metastable $^{2}F_{7/2}$ state and do not interact with the laser beams~\cite{ourpaper_parametricoscillator}. The remaining ion is always positioned at the edge of the ion chain to enable addressing of all the modes of motion.  We use the standard optical pumping to initialize the ion in the $|{\downarrow}\rangle \equiv |S_{1/2}, F=0, m_{F}=0\rangle$ state. Resonance fluorescence technique~\cite{2007PRA} detects the ion in the state $|{\uparrow}\rangle \equiv |S_{1/2}, F=1, m_{F}=0\rangle$. The optical dipole force is applied to the ion in the state $|a\rangle \equiv |S_{1/2}, F=1, m_{F}=+1\rangle$.

\subsection{Experimental sequence}
All experiments commence by preparing the thermal and squeezed thermal motional states while the modes are non-interacting. The energy exchange between modes is switched on for a time $\tau$ by changing their oscillation frequencies. The interaction between the modes is then switched off, and their final state is characterized.

\subsection{State preparation}
At the beginning of every experimental sequence, all nine motional modes are initialized to ground state (residual $\bar{n}_0\leq 0.05$) via Sisyphus cooling~\cite{haljan_2016}, followed by Raman sideband cooling~\cite{monroe_sbc_1995}. It is carried out at detuning $\Delta=\omega_a-\omega_b-\omega_c\approx -2\pi\times40~\mathrm{kHz}$, which is much larger than the coupling rate $\xi$, such that the modes coupling is effectively switched off.

To prepare thermal state, we transfer the ion into the state $|a\rangle$ and excite its motion with modulated optical dipole force. The force is applied by a running optical lattice formed by two linearly polarized beams R1 and R2 (R3 and R2) with orthogonal polarizations (see Fig.~\ref{fig:heat_flow}\,a)~\cite{ourpaper_microwave,ourpaper_parametricoscillator}. The frequency difference between these two beams is set to match the frequency of the target mode while the phase of one of the beams is changed randomly every $100~\mu s$ step. The motional state of the ions undergoes a random walk in phase space, which leads to a thermal state if the number of steps is large enough~\cite{Loudon_book}. 
Typically we apply from 7 to 40 steps for state preparation. The final mean phonon number of the thermal state $\bar{n}$ after $N$ steps is 
\begin{equation}
\bar{n} = \bar{n}_0 + N \bar{m},
\label{eq:thermalstatephononnumber}
\end{equation}
where $\bar{n}_0$ is the mean phonon number after sideband cooling, and $\bar{m}$ is the mean phonon number of a coherent state after applying a single $100\;\mu s$ step to the initial vacuum state (See Section~\ref{sec:therm_state_prep}).

The squeezed thermal state is generated by application of the squeezing operator $\hat{S}(z)=\text{exp}((z^*\hat{a}^2-z\hat{a}^{\dagger 2})/2)$ to a thermal state~\cite{1989squeezedthermalKnight,1992squeezedthermal} where $z=re^{i\theta}$ and $r$ is squeezing parameter. Experimentally, the squeezing operation is realized by applying an optical dipole force produced by an optical lattice running at twice the mode frequency~\cite{1996PRLNonclassicalstate,ourpaper_microwave,ourpaper_parametricoscillator}. The squeezing parameter $r$ is linearly proportional to the duration of this step (see Section~\ref{sec:squeezing_operation}).

\subsection{Energy exchange}
After preparing the motional modes we adjust the offset voltages applied to trap electrodes via low pass filters (LPF). This brings the modes to resonance ($\Delta=0$) with a delay of $25~\mu\mathrm{s}$ which is much smaller than $1/\xi$. The coupling rate is measured to be $\xi=2\pi\times2.64(5)$ kHz for data presented in Fig.~\ref{fig:axial_mode_evol_and_equil} and Fig.~\ref{fig:cooling_cold_mode}\,(a-g), and $\xi=2\pi\times1.89(4)$ kHz for Fig.~\ref{fig:cooling_cold_mode}\,(h-l) and Fig.~\ref{fig:eq_violation}.
After interacting for time $\tau$, the motional modes are decoupled by reverting the detuning back to $\Delta\approx-2\pi\times40\,\mathrm{kHz}$, where the motional states are mapped onto ion internal state for state analysis~\cite{1996PRLNonclassicalstate}. 

\subsection{Motional state detection}
State detection of a mode of interest, after some interaction time $\tau$, can be done by measuring the probability $p_{\uparrow}(\tau)$ to find the detection ion in the ``bright'' internal state $|{\uparrow}\rangle$, after driving a red motional sideband between $|{\downarrow}\rangle$ and $|{\uparrow}\rangle$ with a pulse of fixed duration $t_{\mathrm{rsb}}$.
This probability is dependent on the population distribution $p(n,\tau)$, and has the form :
\begin{equation}
\label{eq:meas_brightness_rsb}
p_{\uparrow}(\tau) = a + \eta \sum_{n=0}^{\infty} p(n,\tau) (1 - \cos(\sqrt{n} \Omega\,t_{\mathrm{rsb}})) / 2,
\end{equation}
where $\Omega$ is the Rabi frequency of the red-sideband, $a$ is the background contribution to the state-detection probability, and $\eta$ is defined as the probability to detect an ion in the state $|{\uparrow}\rangle$ after a $\pi$ pulse on a blue sideband transition 
$|0,\downarrow \rangle \rightarrow |1, \uparrow\rangle$, where the first index corresponds to the motional Fock state.

Typically the population distribution $p(n,\tau)$ is expected to have some analytic time-independent form. Then one can compute the inverse function $p_{\uparrow}^{-1}(\tau)$ that links the measured ion brightness $p_{\uparrow exp}(\tau)$ to mean phonon number $\bar{n}(\tau)$. However, during the interaction, the states evolve away from their initial thermal (or squeezed thermal) population distribution. Then $p(n,\tau)$ is not known a-priori. We therefore compute for each given set of initial states $(\bar{n}_h^{(in)},\bar{n}_w^{(in)},\bar{n}_c^{(in)})$ the expected spin-flip probability $p_{\uparrow th}(\tau)$ and the mean phonon number $\bar{n}_{th}(\tau)$ during the state evolution, by numerically solving \eqref{eq:interaction}. In order to obtain the best estimate for the experimental mean phonon numbers, we determine $\bar{n}_{exp}(\tau)$ from the experimentally measured spin-flip probability $p_{\uparrow exp}(\tau)$ using:
\begin{equation}
\label{eq:nbar_scaling}
\bar{n}_{exp}(\tau) \approx \bar{n}_{th}(\tau) + \frac{\partial \bar{n}_{th}(\tau)}{\partial p_{\uparrow th}(\tau)}[p_{\uparrow exp}(\tau) - p_{\uparrow th}(\tau)].
\end{equation}
The partial derivative in~\eqref{eq:nbar_scaling} is approximated by 
\[\frac{\partial \bar{n}_{th}(\tau)}{\partial p_{\uparrow th}(\tau)} \approx \frac{\bar{n}_{th}(\tau;\bar{n}_i^{(in)} + \delta) - \bar{n}_{th}(\tau; \bar{n}_i^{(in)} - \delta)}{p_{\uparrow th}(\tau; \bar{n}_i^{(in)} + \delta) - p_{\uparrow th}(\tau; \bar{n}_i^{(in)} - \delta)},\]
where $\bar{n}_i^{(in)}$ is the initial population of the mode and $\delta$ is a small but finite deviation from $\bar{n}_i^{(in)}$. For example, if the cold mode is the mode of interest, the numerical simulations of $\bar{n}_c(\tau)$ would be carried out for $\bar{n}_c^{(in)}\pm\delta$ with fixed $\bar{n}_h^{(in)}$ and $\bar{n}_w^{(in)}$.

\subsection{Numerical simulation}
The interaction Hamiltonian~\eqref{eq:interaction} couples Fock states of the form
\begin{equation}
	\{|n_h, N-n_h,M-n_h \rangle :\ 0\leq n_h\leq min(N,M)\}
\end{equation}
with fixed integers $N$ and $M$. This basis spans a finite-dimensional Hilbert space. The evolution of the three-mode state is then computed by diagonalizing the Hamiltonian in each of the contributing subspaces, up to a cutoff for both $N$ and $M$. For all the simulations presented in this paper, the cutoff has been chosen to ignore terms in the density matrix smaller than $10^{-4}$.
We also implemented an incoherent version of the interaction by integrating the master equation $\partial_t \rho = -\xi_{\rm in} [\hat{H}, [ \hat{H}, \rho]]$, which describes an exponential decay of coherences in the eigenbasis of the Hamiltonian at the rate $2\xi_{\rm in}$. The fully decohered state represents the long time average phonon numbers of the coherently evolving state. However, the incoherent model does not reproduce the single-shot cooling behavior.

\subsection{Calibration of the state preparation\label{sec:state_calibration}}
The reproducible operation of the refrigerator requires careful calibration of the initial mean phonon numbers of the motional mode. Below we describe the procedures for preparation of thermal and squeezed thermal states and methods employed for the calibration. 
\subsubsection{Reconstruction of the phonon number distribution for the calibration procedure}
To reconstruct the phonon number distribution, we drive the blue sideband transition and measure the temporal evolution of internal state of the ion. The data is then fitted to the function of the form~\cite{1996PRLNonclassicalstate}
\begin{equation}
\label{eq:general_fit}
p_{\uparrow}(t)=\frac{a}{2}\left( 1-\sum_{n=0}p(n)\,\mathrm{cos}\left(\Omega_{n,n+1}\,t\right)e^{-\gamma_n\,t}\right)+b,
\end{equation}
where $p(n)$ is the expected population distribution for the target quantum state of the mode, 
$\Omega_{n,n+1}=\sqrt{n+1}\,\Omega_{0,1}$ is the state dependent Rabi frequency and $\gamma_n=\sqrt{n+1}\,\gamma_0$ is the decoherence rate~\cite{1996PRLNonclassicalstate}. Parameters $a$, $b$ are introduced to account for the imperfect phonon detection efficiency and state detection background. The results of the fit are used to calibrate the preparation procedure for all the states used in the experiment. 

\subsubsection{Thermal state \label{sec:therm_state_prep}}
\begin{figure}
\centering
\includegraphics[width=\columnwidth]{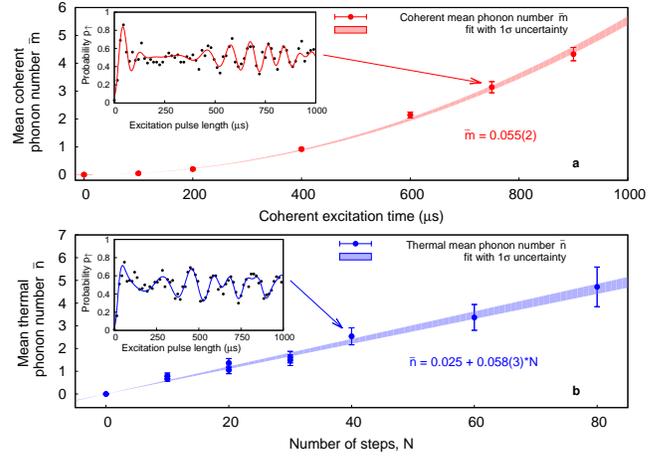}
\caption{\label{fig:coh_and_thermal}
{\bf Thermal state preparation. a.} Mean phonon number of the coherent states as a function of excitation time. It is fitted to the function of the form $\bar{m}_t=n_0 + \beta\times t^2$ to extract $\bar{m}=\beta\times(100\,\mu s)^2$ in Eq.~\eqref{eq:thermalstatephononnumberSI}. Here $n_0=0.025$ being the independently measured residual mean phonon number after sideband cooling. {\bf b.} Mean phonon number dependence on the number of preparation steps. The linear fit yields the mean phonon number of the thermal state according to equation~\eqref{eq:thermalstatephononnumberSI}. The insets show the internal state evolution versus blue sideband excitation pulse length for coherent ({\bf a}) and thermal ({\bf b}) states. The fits to equation~\eqref{eq:general_fit} with Poissonian~\eqref{eq:poisson} and thermal~\eqref{eq:thermal} distributions yield the mean phonon numbers.}
\end{figure}

During the preparation of a thermal state, each step displaces the ions motional state by length $\alpha$ in a random direction in phase space. After $N$ steps, the expected displacement from the origin is $\sqrt{N}\,\alpha$, which corresponds to $\bar{n}=N|\alpha|^2$ phonons, where $|\alpha|^2=\bar{m}$ is the mean phonon number of coherent state after one step~\cite{Loudon_book}. Adding the initial phonon population $\bar{n}_0$ after the imperfect sideband cooling~\cite{wineland_review} then gives 
\begin{equation}
\bar{n} = \bar{n}_0 + N \bar{m}.
\label{eq:thermalstatephononnumberSI}
\end{equation}
To experimentally verify this equation, we extract $\bar{m}$ and $\bar{n}$ by fitting the temporal evolution of coherent states and thermal states to Eq.~\eqref{eq:general_fit}, where the expected  population distributions are
\begin{equation}
\label{eq:poisson}
p(n)=\frac{\bar{m}^n\,e^{-\bar{m}}}{n!},
\end{equation}
and  
\begin{equation}
\label{eq:thermal}
p(n)=\frac{\bar{n}^n}{(\bar{n}+1)^{n+1}},
\end{equation}
respectively. The results are shown in 
Fig.~\ref{fig:coh_and_thermal}, where the hot mode excitation is taken as an example. They are consistent with the prediction of Eq.~\eqref{eq:thermalstatephononnumberSI}.

\subsubsection{Squeezing operation\label{sec:squeezing_operation}}
\begin{figure}
\centering
\includegraphics[width=\columnwidth]{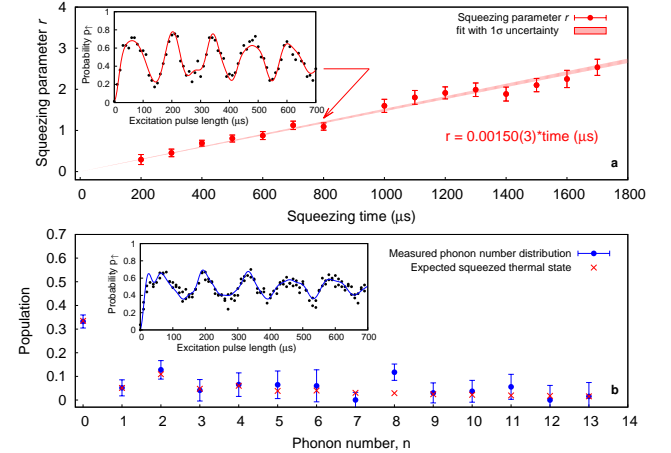}
\caption{\label{fig:sqz_vac_and_thermal}
{\bf Squeezed vacuum and squeezed thermal state preparation. a.} Squeezing parameter $r$ in the work mode as a function of squeezing operation time $t$ applied to the vacuum state. The red dots are the experimental data and the red line is a fit using $r=\rho\times t$. The inset shows the internal state evolution corresponding to squeezing time of $800 \mu s$. The red curve is the fit with equation~\eqref{eq:sqz_vac} which yields squeezing parameter $r=1.09(9)$.  
{\bf b.} The phonon number population distribution of the squeezed thermal state. 
The distribution reconstructed (blue dots) from the fit to the experimental data (inset) where all the $p(n)$ are allowed to vary independently is compared to the distribution calculated from the expected values for $r=1.2$ and $\bar{n}=0.77$ (red crosses).}
\end{figure}

The squeezing operation on the work mode can be independently calibrated using squeezed vacuum state, which is prepared by starting from the vacuum state and then applying the optical dipole force at twice the trap frequency for some time. The expected population distribution of the squeezed vacuum state is restricted to even number states
\begin{equation}
\label{eq:sqz_vac}
p(2n)=\frac{(2n)!\,\mathrm{sech}r\,\mathrm{tanh}^{2n}r}{(2^n\,n!)^2}.
\end{equation}
The squeezing parameter $r$ is linearly proportional to the duration of the applied force~\cite{1996PRLNonclassicalstate}, as shown in Fig.~\ref{fig:sqz_vac_and_thermal}\,\textbf{a}.

To demonstrate the coherence of the squeezing operation, we apply a second squeezing pulse with the same pulse amplitude as the first one but $180^{\circ}$ out of phase. We verify that this pulse brings the ions back to ground state with probability of more than $80\%$ for $r=1.2$.

\subsubsection{Squeezed thermal state}
The population distribution of squeezed thermal state with squeezing parameter $r$ and initial phonon number $\bar{n}$ is~\cite{1989squeezedthermalKnight,1992squeezedthermal}
\begin{equation}
p(n) =\sum_{m=0}\frac{\bar{n}^m}{(\bar{n}+1)^{m+1}} D_n^{SN}(m, r),
\label{eq:squeezedthermal}
\end{equation}
where 

\begin{widetext}
\[D_n^{SN}(m, r)=
\left\{
	\begin{array}{lllll}
		\ddfrac{n!\,m!}{((m/2)!\,(n/2)!)^2}\ddfrac{1}{\text{cosh}\,r}(\ddfrac{\text{tanh}\,r}{2})^{(m+n)} \times & \\
        \qquad \qquad (_2F_1(-\ddfrac{n}{2},-\ddfrac{m}{2};\ddfrac{1}{2};-\ddfrac{1}{(\text{sinh}\,r)^2}))^2  & \mbox{for m \& n even} \\[1em]
		\ddfrac{n!m!}{(((m-1)/2)!((n-1)/2)!)^2}\ddfrac{1}{(\text{cosh}\,r)^3}(\ddfrac{\text{tanh}\,r}{2})^{(m+n-2)} \times & \\
        \qquad \qquad (_2F_1(-\ddfrac{n-1}{2},-\ddfrac{m-1}{2};\ddfrac{3}{2};-\ddfrac{1}{(\text{sinh}\,r)^2}))^2 & \mbox{for m \& n odd} \\[1em]
        0 & \mbox{otherwise}
	\end{array}
\right.\]
\end{widetext}
is the population distribution of the squeezed number state and $_2F_1$ is the Gaussian hypergeometric function.

The squeezed thermal state is prepared by applying the squeezing operation to some thermal state. We then fit the temporal evolution of the internal state to Eq.~\eqref{eq:general_fit}, while allowing the fit parameters $p(n)$ for $n$ ranging from 0 to 13 to vary, constrained only by $\sum_n p(n) = 1$. The extracted population distribution is compared to the one predicted by Eq.~\eqref{eq:squeezedthermal} with the expected values of $r$ and $\bar{n}$.

\subsubsection{Calibration of experimental initial conditions.}
\begin{figure}
\centering
\includegraphics[width=\columnwidth]{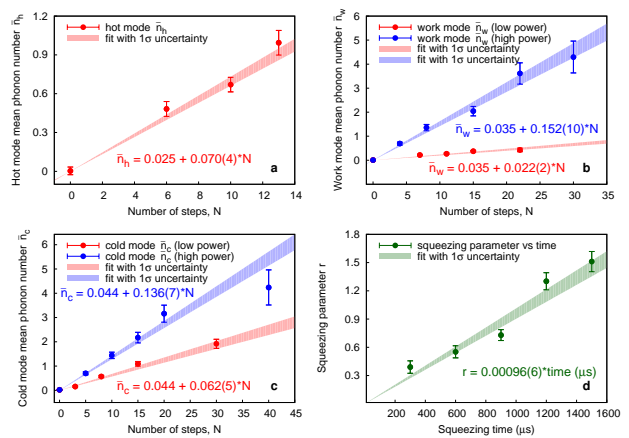}
\caption{\label{fig:calibration curves}
{\bf Calibration of the thermal and squeezed state parameters for all modes. a.} Dependence of $\bar{n}_h$ in the hot mode on the number of random steps at which the lattice was applied. {\bf b.} Same for work mode $\bar{n}_w$ for two different lattice depths. The different depths were employed to cover wider range of initial temperatures. {\bf c.} Same for cold mode. The offset values in the fit functions represent independently measured residual mean phonon numbers after sideband cooling. {\bf d.} Squeezing parameter $r$ of the work mode versus application time of the optical lattice running at $2\omega_w$ and trap being set to lower frequency (see Methods). The values of $r$ were extracted by fitting squeezed thermal state evolution with equation~\eqref{eq:squeezedthermal} with both $\bar{n}$ and $r$ being fit parameters.}
\end{figure}

In Fig.~\ref{fig:calibration curves} we show the results of the calibration of the thermal mean phonon numbers and squeezing parameter.
\begin{table*}
\centering
\begin{tabular}{ |M{5.2cm}|M{7.5cm}|M{3.2cm}|}
\hline
\multicolumn{3}{|c|}{Figure 2 (a)}\\
\hhline{|===|}
Number of steps $(h,w,c)$ & $\bar{n}_i^{in}(h,w,c)$ & $\bar{n}_h^{ss}$\\
\hline
$(9, 29\mathrm{H}, 7\mathrm{L})$ & $(0.66(4), 4.44(29), 0.48(4))$ & $0.46(1)$ \\
\hline
$(9, 29\mathrm{H}, 14\mathrm{L})$ & $(0.66(4), 4.44(29), 0.91(7))$ & $0.60(2)$\\
\hline
$(9, 29\mathrm{H}, 10\mathrm{H})$ & $(0.66(4), 4.44(29), 1.40(7))$ & $0.715(6)$\\
\hline
$(9, 29\mathrm{H}, 13\mathrm{H})$ & $(0.66(4), 4.44(29), 1.81(9))$ & $0.81(2)$ \\
\hline
$(9, 29\mathrm{H}, 17\mathrm{H})$ & $(0.66(4), 4.44(29), 2.36(12))$ & $0.91(4)$ \\
\hline
$(9, 29\mathrm{H}, 20\mathrm{H})$ & $(0.66(4), 4.44(29), 2.76(14))$ & $1.07(5)$ \\
\hhline{|===|}
\multicolumn{3}{|c|}{Figure 2 (b)}\\
\hhline{|===|}
Number of steps $(h,w,c)$ & $\bar{n}_i^{in}(h,w,c)$ & $\bar{n}_h^{ss}$\\
\hline
$(9, 16\mathrm{H}, 7\mathrm{L})$ & $(0.66(4), 2.47(0.16), 0.48(4))$ & $0.50(2)$ \\
\hline
$(9, 16\mathrm{H}, 14\mathrm{L})$ & $(0.66(4), 2.47(0.16), 0.91(7))$ & $0.63(1)$\\
\hline
$(9, 16\mathrm{H}, 10\mathrm{H})$ & $(0.66(4), 2.47(0.16), 1.40(7))$ & $0.64(2)$ \\
\hline
$(9, 16\mathrm{H}, 13\mathrm{H})$ & $(0.66(4), 2.47(0.16), 1.81(9))$ & $0.72(3)$\\
\hline
$(9, 16\mathrm{H}, 17\mathrm{H})$ & $(0.66(4), 2.47(0.16), 2.36(12))$ & $0.73(3)$ \\
\hline
$(9, 16\mathrm{H}, 20\mathrm{H})$ & $(0.66(4), 2.47(0.16), 2.76(14))$ & $0.75(3)$\\
\hhline{|===|}
\multicolumn{3}{|c|}{Figure 2 (c)}\\
\hhline{|===|}
Number of steps $(h,w,c)$ & $\bar{n}_i^{in}(h,w,c)$ & $\bar{n}_h^{ss}$\\
\hline
$(8, 10, 8\mathrm{L})$ & $(0.64(4), 1.82(8), 0.64(4))$ & $0.43(1)$ \\
\hline
$(8, 10, 15\mathrm{L})$ & $(0.64(4), 1.82(8), 1.17(8))$ & $0.54(2)$\\
\hline
$(8, 10, 23\mathrm{L})$ & $(0.64(4), 1.82(8), 1.77(12))$ & $0.565(6)$\\
\hline
$(8, 10, 29\mathrm{L})$ & $(0.64(4), 1.82(8), 2.22(15))$ & $0.68(2)$ \\
\hline
$(8, 10, 35\mathrm{L})$ &$(0.64(4), 1.82(8), 2.67(18))$ & $0.697(8)$\\
\hhline{|===|}
\multicolumn{3}{|c|}{Figure 2 (d)}\\
\hhline{|===|}
Number of steps $(h,w,c)$ & $\bar{n}_i^{in}(h,w,c)$ & $\bar{n}_h^{ss}$\\
\hline
$(9, 7\mathrm{H}, 7\mathrm{L})$ & $(0.66(4), 1.10(7), 0.48(4))$ & $0.49(1)$ \\
\hline
$(9, 7\mathrm{H}, 14\mathrm{L})$ & $(0.66(4), 1.10(7), 0.91(7))$ & $0.51(1)$\\
\hline
$(9, 7\mathrm{H}, 10\mathrm{H})$ & $(0.66(4), 1.10(7), 1.40(7))$ & $0.59(2)$ \\
\hline
$(9, 7\mathrm{H}, 13\mathrm{H})$ & $(0.66(4), 1.10(7), 1.81(9))$ & $0.64(2)$\\
\hline
$(9, 7\mathrm{H}, 17\mathrm{H})$ & $(0.66(4), 1.10(7), 2.36(12))$ & $0.62(2)$ \\
\hline
$(9, 7\mathrm{H}, 20\mathrm{H})$ & $(0.66(4), 1.10(7), 2.76(14))$ & $0.74(1)$\\
\hline
\end{tabular}
\caption{Measured steady state values for the presented data. The left column shows the number of phase modulating steps for thermal state preparation in each mode. ``H'' or ``L'' letters indicate high or low power of the optical lattice. Initial mean phonon numbers obtained from calibration measurements are shown in the middle column. The measured steady state values are presented in the right column.}
\end{table*} 

\begin{table*}
\centering
\begin{tabular}{ |M{5.2cm}|M{7.5cm}|M{3.2cm}|}
\hline
\multicolumn{3}{|c|}{Figure 3 (a-f)}\\
\hhline{|===|}
Number of steps $(h,w,c)$ & $\bar{n}_i^{in}(h,w,c)$ & $\bar{n}_c^{ss}$\\
\hline
$(9, 29\mathrm{H}, 19\mathrm{H})$ & $(0.66(4), 4.44(29), 2.63(13))$ & $2.11(3)$ \\
\hline
$(9, 14\mathrm{H}, 19\mathrm{H})$ & $(0.66(4), 2.16(14), 2.63(13))$ & $2.58(5)$ \\
\hline
$(9, 7\mathrm{H}, 19\mathrm{H})$ & $(0.66(4), 1.10(7), 2.63(13))$ & $2.53(2)$\\
\hline
$(9, 29\mathrm{L}, 19\mathrm{H})$ & $(0.66(4), 0.67(6), 2.63(13))$ & $2.61(4)$ \\
\hline
$(9, 15\mathrm{L}, 19\mathrm{H})$ & $(0.66(4), 0.37(3), 2.63(13))$ & $2.70(7)$\\
\hline
$(9, 7\mathrm{L}, 19\mathrm{H})$ & $(0.66(4), 0.19(1), 2.63(13))$ & $2.92(8)$ \\
\hhline{|===|}
\multicolumn{3}{|c|}{Figure 3 (h-k)} \\
\hhline{|===|}
Number of steps $(h,w, {\bf squeezing\,time}, c)$ & $\bar{n}_i^{in}(h,w,{\bf r},c)$ & $\bar{n}_c^{ss}$ \\
\hline
$(9, 8, \bm{1400}\, \mu s, 19\mathrm{H})$ & $(0.47(6), 0.50(5), \bm{1.34(8)}, 2.60(3))$ & $2.26(6)$ \\
\hline
$(9, 8, \bm{1200}\, \mu s, 19\mathrm{H})$ & $(0.52(6), 0.50(5), \bm{1.15(7)}, 2.72(4))$ & $2.46(4)$\\
\hline
$(9, 8, \bm{800}\, \mu s, 19\mathrm{H})$ & $(0.52(6), 0.50(5), \bm{0.77(4)}, 2.81(4))$ & $2.76(3)$\\
\hline
$(9, 8, \bm{0}\, \mu s, 19\mathrm{H})$ & $(0.46(6), 0.50(5), \bm{0}, 3.01(4))$ & $3.26(6)$ \\
\hline
\end{tabular}
\newline\newline
\caption{Measured steady state values for the presented data. The left column shows the number of phase modulating steps for thermal state preparation in each mode and the squeezing time used to prepare a squeezed thermal state. Initial mean phonon numbers obtained either from calibration or direct measurements are shown in the middle column. The measured steady state values are presented in the right column.}
\end{table*}

The calibration presented in Fig.~\ref{fig:calibration curves}\,(a-c) was used to obtain the initial mean phonon numbers shown in Fig.~\ref{fig:axial_mode_evol_and_equil}\,(a,b,d) and~\ref{fig:cooling_cold_mode}\,(a-f).~\footnote{Data presented at Fig.~\ref{fig:axial_mode_evol_and_equil}\,c was taken on different date and used different calibration, which was obtained in exactly the same way. The calibration equations for this data are: $\bar{n}_h=0.025 + 0.075(5)\cdot\mathrm{step}$ for the hot mode, $\bar{n}_w=0.035+0.178(8)\cdot\mathrm{step}$ for the work mode and $\bar{n}_c^{L}=0.044+0.075(5)\cdot\mathrm{step}$ $\left(\bar{n}_c^{H}=0.044+0.172(8)\cdot\mathrm{step}\right)$ for the cold mode with low (high) power in the lattice.} Due to air conditioning failure causing the technical disruption to the experiment the calibration was made obsolete and the trap frequency had to be lowered. To make sure that such disruptions will not alter calibration and data acquisition again, in subsequent experiments we determine the information about the initial state just before we start the measurements. For the experiments performed with lower trap frequencies, for which results are shown in Fig.~\ref{fig:cooling_cold_mode}~(h-k) and Fig.~\ref{fig:eq_violation} the initial mean phonon numbers of the cold mode were obtained directly from the measured $p_{\uparrow}$ at $\tau = 0$ with the help of~\eqref{eq:meas_brightness_rsb}. For the data taken with squeezed work mode in Fig.~\ref{fig:cooling_cold_mode}~(h-k), the initial values for the hot mode were measured once using same number of steps at the beginning of every evolution branch. The work mode initial thermal mean phonon number was measured independently and the values of $r$ were obtained from the calibration shown in Fig.~\ref{fig:calibration curves}\,(d).
For the data presented in Fig.~\ref{fig:eq_violation} the work mode initial mean phonon numbers were measured as described in section~\ref{sec:therm_state_prep} for every individual point while the hot mode initial mean phonon number was measured once after all the points were taken following the same procedure.

\subsection{Steady state values}
To compensate for slow systematic drifts of the initial mean phonon numbers we plot the measured mean phonon numbers of the refrigerator modes relative to the steady state values. This approach does not affect any of our conclusions, but provides better visualization of data. In the Tables we show the measured steady state values as well as the initial conditions for all the relevant experiments. The steady state values $\bar{n}_{ss}$ that were used to generate Fig.~\ref{fig:axial_mode_evol_and_equil},~\ref{fig:cooling_cold_mode}~(a-f), were obtained by averaging the measured $\bar{n}(\tau)$ for $\tau > 240\,\mu$s. For Fig.~\ref{fig:cooling_cold_mode}\,(h-k)the averaging was performed  for $\tau > 600\,\mu$s.

\begin{acknowledgments}
 We acknowledge discussions with Alex Kuzmich, Atac Imamoglu and Mark Mitchison. 
 This research is supported by the National Research Foundation, Prime Minister’s Office, Singapore and the Ministry of Education, Singapore under the Research Centres of Excellence programme and Education Academic Research Fund Tier 2 (Grant No. MOE2016-T2-1-141).
\end{acknowledgments}


\end{document}